\documentclass[12pt,aps,prd,preprint,tightenlines,
   showpacs,nofootinbib]{revtex4}
\newcommand{\PRE}[1]{{#1}} 

\usepackage{bm} \usepackage{epsfig}
\newcommand{\eqref}[1]{Eq.~(\ref{#1})}

\newcommand{\secref}[1]{Sec.~\ref{sec:#1}}

\newcommand{\figref}[1]{Fig.~\ref{fig:#1}}

\begin{document}

\preprint{UCI-TR-2007-31}

\title{
\PRE{\vspace*{1.5in}}
Determining  Spins of Metastable Sleptons   at the Large Hadron Collider

\PRE{\vspace*{0.3in}} }

\author{Arvind Rajaraman}
\affiliation{Department of Physics and Astronomy, University of
California, Irvine, CA 92697, USA \PRE{\vspace*{.5in}} }
\author{Bryan T.~Smith%
\PRE{\vspace*{.2in}} } \affiliation{Department of Physics and
Astronomy, University of California, Irvine, CA 92697, USA
\PRE{\vspace*{.5in}} }

\begin{abstract}
\PRE{\vspace*{.3in}}
We investigate models in  minimal supergravity parameter space
which contain metastable sleptons. We find the luminosity required
to determine the slepton spin in these scenarios,
and apply our analysis to two benchmark models.
We show that the spin of the  slepton in one of
the benchmark models can be determined with less than 30 fb$^{-1}$ of data,
while the slepton spin in the other model can be determined with roughly
40 fb$^{-1}$ of data.  We show how our analysis can be applied to other models, and
 give an estimate of the luminosity needed for
a spin determination as a function of slepton mass. This analysis can be
used to distinguish supersymmetry and extra dimensions.
\end{abstract}


\maketitle

\section{Introduction}
\label{sec:introduction}

In the near future, the LHC will begin taking data, and there are
high expectations that there will be a discovery of new physics
beyond the Standard Model. It
will be a further challenge to isolate the nature of the new physics
from the many possible scenarios. In particular,
supersymmetric models can appear very similar to models of extra dimensions, and
it is often difficult to tell them apart~\cite{Cheng:2002ab}. In fact, the best
way to distinguish these
two classes of models is by measuring the spins of the new particles.

Motivated by
an effort to
distinguish between supersymmetry and extra dimensions, there has been
a great deal of   work done on measuring the spins of
beyond the standard model~(BSM) particles.  These studies
have mostly
focused on the case of missing energy
signals~\cite{Barr:2004ze,Smillie:2005ar,Datta:2005zs,Barr:2005dz,Alves:2006df,Athanasiou:2006ef,Wang:2006hk,Alves:2007xt}.
These works considered cascades of heavier particles decaying down to a neutral
LSP, looking at the kinematic distribution of jets and leptons.

In this note we shall discuss a different class of models,
in which the signals do not come from missing energy, but
instead from  metastable charged particles.
 We shall
 focus on the possibility of measuring the spins of these new particles, and thereby
 distinguishing supersymmetry from extra dimensions.

The models we discuss
are part of  minimal supergravity (mSugra)~\cite{Chamseddine:1982jx} parameter space.
In these models, the gravitino is the lightest superpartner (LSP),
  while a charged slepton is the next-to-lightest superpartner (NLSP).
Such models
are typically considered to be ruled out.  If the gravitino is ignored (as is typical in
supersymmetry analyses),
the
slepton is stable. It then
appears as an absolutely stable charged massive particle (CHAMP).
There are very strong bounds on CHAMP masses~\cite{Smith:1979rz,Dimopoulos:1989hk,Pospelov:2006sc,Cyburt:2006uv}. This
constraint removes a large region of mSugra parameter space.

However, as pointed out in~\cite{Feng:2005ba}, CHAMP constraints are relaxed in the presence of
 the gravitino~\cite{Feng:2003xh,%
Ellis:2003dn,Feng:2004zu,Wang:2004ib,Ellis:2004bx,%
Roszkowski:2004jd,Brandenburg:2005he,Cerdeno:2005eu,Kaplinghat:2005sy,Cembranos:2005us,Jedamzik:2005sx,%
Sigurdson:2003vy,Profumo:2004qt,Steffen:2006hw,Steffen:2006wx,Pradler:2006qh}.
The charged slepton  will decay to the gravitino in a
short time period. The
current dark matter will then be entirely composed of gravitinos, and
CHAMP searches will not put any constraints on this scenario.
  Much of the parameter space in the
region with slepton NLSPs is consistent with all current limits.

The first comprehensive analysis of the mSugra parameter space with
metastable sleptons
was performed in a previous paper~\cite{Rajaraman:2006mr}, showing that
a large portion of this parameter space can be probed at the first high
energy run of the Large Hadron Collider~(LHC) with no more than 1-4 fb$^{-1}$ of data.  Some
regions of parameter space  may even have a ``Day 1''
discovery of the meta-stable charged sleptons.

In this paper, we shall extend these analyses to consider the
possibility of measuring the
spin of the NLSP sleptons.  We will consider a direct
measurement of the spin of the slepton NLSP from Drell-Yan production.  The
angular and energy distribution of Drell-Yan produced scalars and
fermions are well known, and hence the spin can be measured from the
kinematic data. The analysis will be applied to the extended mSugra benchmark models,
Model A and Model B, that were defined in~\cite{Feng:2005ba}.   It will be shown
that the spin determination is possible for both models with about 40 fb$^{-1}$
of data.


\begin{figure}
\resizebox{6.0in}{!}{
\includegraphics{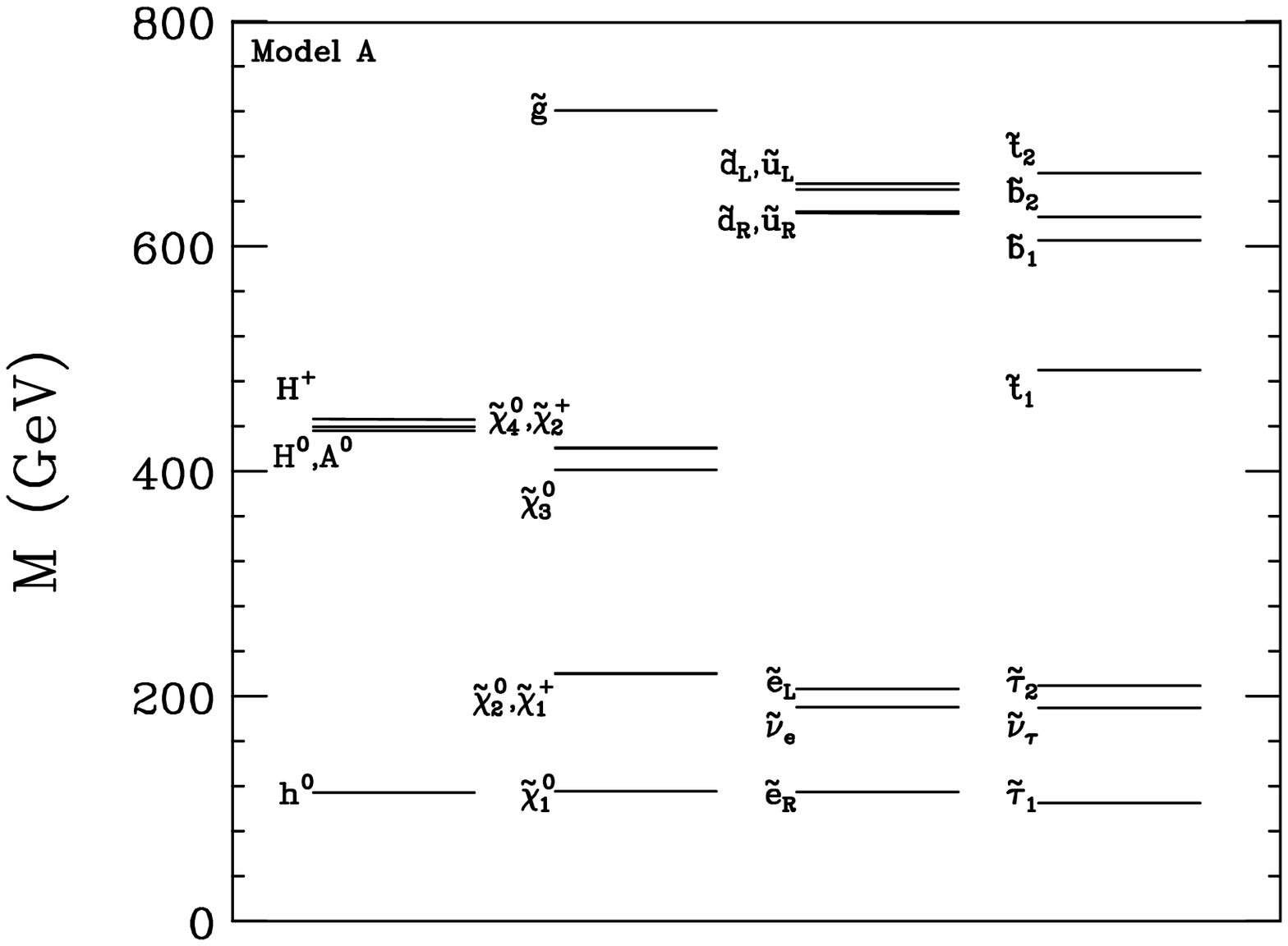} \qquad
\includegraphics{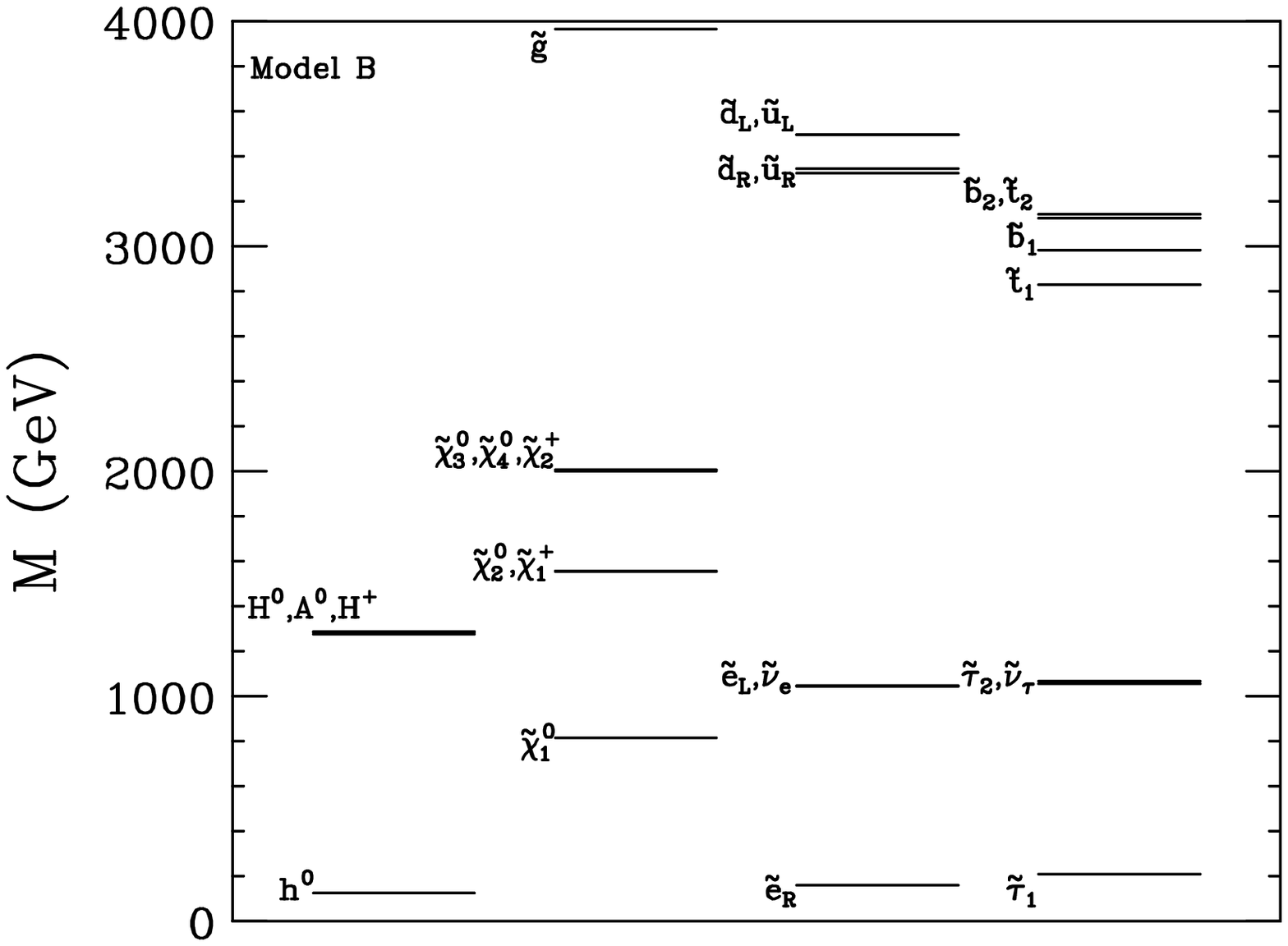}
}
\caption{The SUSY mass spectrum for Model A~(Left) and Model B~(Right).\label{fig:BenchMass}}
\end{figure}

The  structure of this paper is the following. A detailed
analysis of the mass measurement is performed in~\secref{mass}.  Cuts
for the spin measurement are developed in~\secref{cuts}. The
analysis to determine the spin is presented in~\secref{spin},
along with more general estimates for the prospect
of the slepton spin determination as a function of the slepton
mass. \secref{conclusion} contains
a final discussion of our results.

\section{Mass Measurement}
\label{sec:mass}

To find the angular distribution in the center of mass
frame it is necessary to boost back Drell-Yan produced sleptons from
the lab frame.  The boost requires a knowledge of the mass of the sleptons.
  Knowledge of the mass also allows a
 separation of background muons from the signal~\cite{Ellis:2006vu}.
 We therefore begin with a discussion of the mass measurement.

 The mass spectrum of
the two benchmark models can be seen in~\figref{BenchMass}, where the left panel is
the spectrum for Model A, and the right panel is the spectrum for
Model B. The
discovery potential for these models was
studied in~\cite{Rajaraman:2006mr}.     Discovery can be achieved at the LHC with 5.6 pb$^{-1}$ for
Model A and 720 pb$^{-1}$ for Model B; both models are well within the reach
of the LHC first physics run.  The total SUSY cross sections for the
two models are 18 pb for Model A and 43 fb for Model B.

Once the meta-stable charged particles have been
discovered, a mass measurement can be made with a few low velocity
events.   The resolution for the momentum and velocity
measurements are
~\cite{Polesello:1999aa,Ellis:2006vu},
\begin{equation}
\sigma\left(p\right)~(\textrm{GeV}) = 0.000118 p^2 + 0.0002 \sqrt{p^2 +
m^2_{\tilde{\tau}}} + 0.89
\label{pres}
\end{equation}
\begin{equation}
\sigma\left(\beta\right) = 0.028 \beta^2.
\label{bres}
\end{equation}
The low velocity and low momentum events have better detector
resolution, thus having a higher resolution on the mass.

\begin{figure}
\resizebox{6.0in}{!}{
\includegraphics{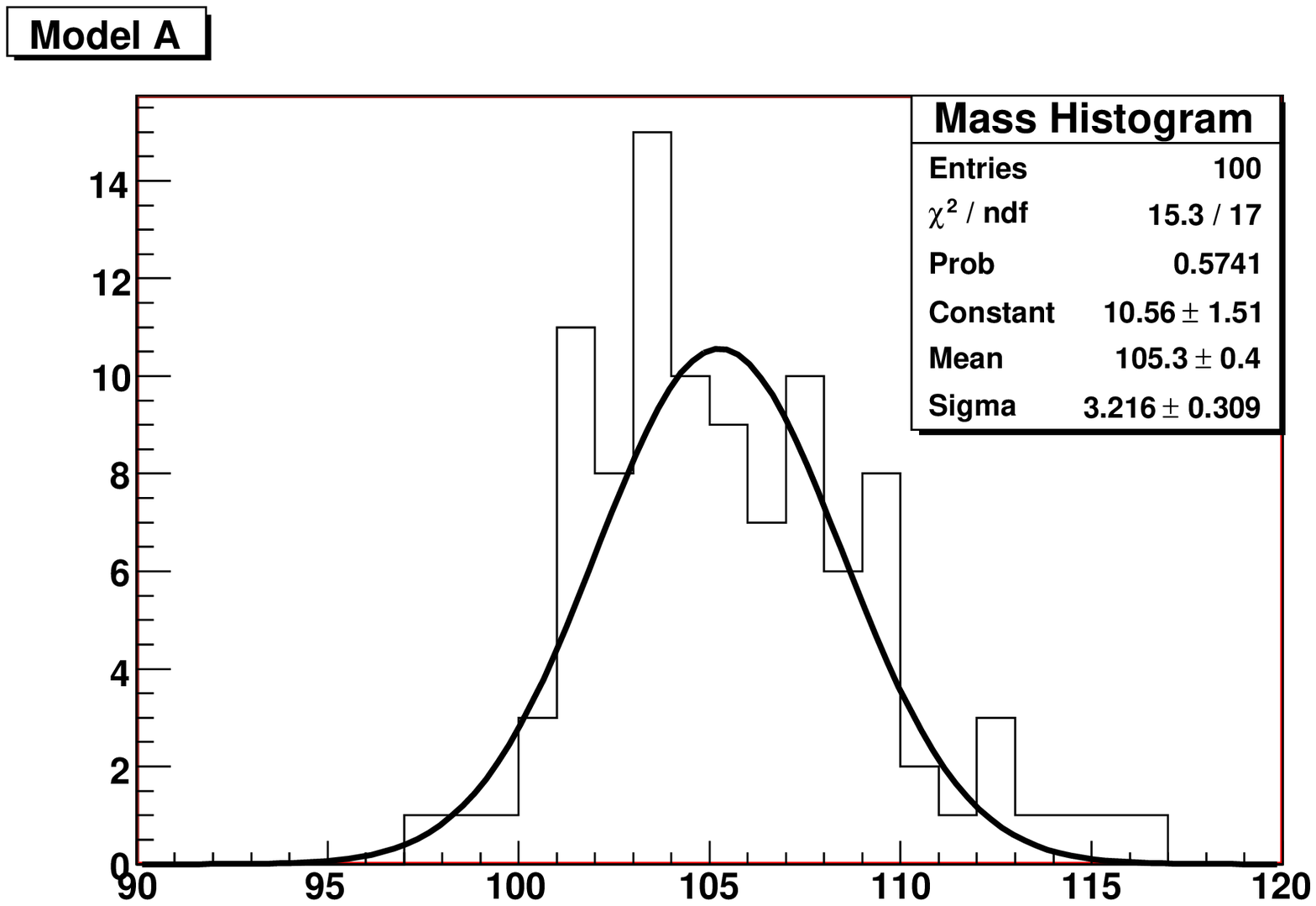} \qquad
\includegraphics{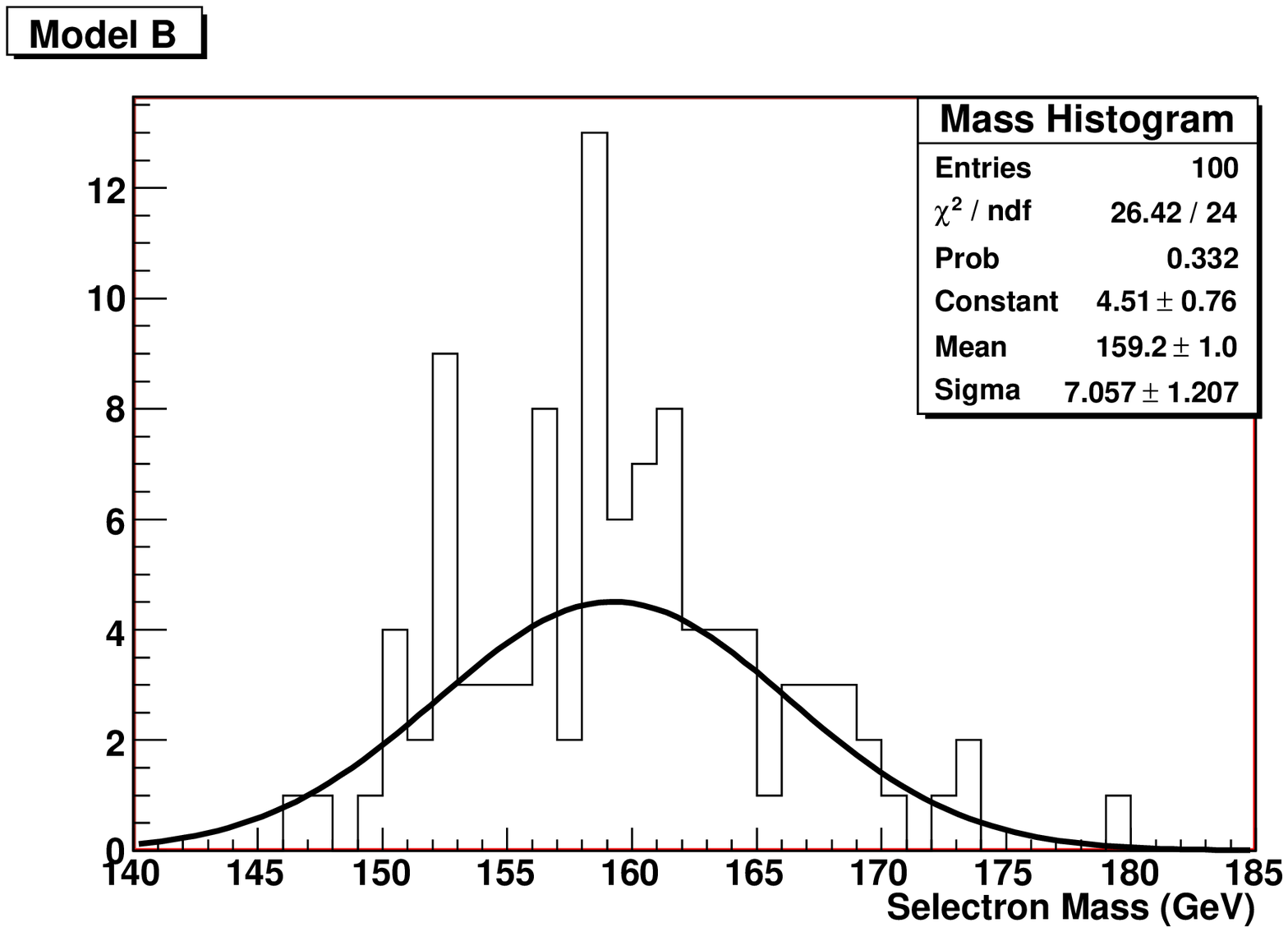}
} \caption{The histogram of $\frac{p_{meas}}{\beta\gamma_{meas}}$ for
  a 105 GeV stau~(left) and a 159 GeV selectron~(right).
\label{fig:mass}}
\end{figure}

   A  mass measurement to better than 5\% is obtainable with as few as 100 events by
taking the ratio of the measured momentum to the measured
$\beta\gamma$.
The mass measurement
 accuracy is illustrated in 
 ~\figref{mass} where the mass of the 105.4 GeV
NLSP stau from Model A is measured in the left panel and the 159.5 GeV
NLSP selectron from Model B is measured in the right panel.
The masses are measured to within a GeV with 100 events, and the mass resolution for
the slower velocity sleptons is a few GeV,
matching~\cite{Polesello:1999aa}.  The luminosity needed for the mass
measurements are 40 pb$^{-1}$ for the stau of Model A and 15 fb$^{-1}$ for the
right handed selectron of Model B, with the use of the following cuts:

{\it a. A slepton candidate:} $M_{meas} =
\frac{\beta\gamma_{meas}}{p_{meas}} > $100 GeV

{\it b. Rapidity cuts:}  $|\eta| < 2.4 $.

 {\it c. Velocity cuts:}  $0.67 < \beta < 0.71$

 {\it d. Transverse Momentum cut:} $p_T > 20$ GeV.

These cuts do not requires that the sleptons are from
Drell-Yan production or that both sleptons are observed in each event.
They do require that at least one slepton
does trigger the detector.  The luminosities quoted above
are an overestimate as a result of this requirement.

\section{Cuts}
\label{sec:cuts}
Drell-Yan events must be isolated from both muons and cascade decays.
 Muon separation can be obtained in a number
of ways.  
In a previous
paper~\cite{Rajaraman:2006mr}, it was shown that
the requirement of a time delay can effectively remove muons from our
signal.  For this paper we will
require a time delay greater than 1ns for the observed sleptons.
Both particles are also required to have a measured velocity and
momentum that are consistent with the 
measured NLSP slepton
mass.  These cuts remove both the SM muon and SUSY muon
background, while maximizing the acceptance of our signal.

The cascade decays are removed by requiring events to look like
Drell-Yan events.  Hence the events should have  two meta-stable
charged particles candidates, without isolated or high energy leptons,
without high energy jets, and without missing transverse energy.
The following set of cuts are adopted to accomplish this:

{\it a. Two slepton candidates:} Two particles with $M_{meas}$ consistent with $M_{slepton}$

{\it b. Rapidity cuts:}  Both candidates with $|\eta| < 2.4 $

 {\it c. Time Delay Cut:} Both candidates with $\Delta t > 1~\rm{ns}$

 {\it d. Transverse Momentum cut:} Both candidates with $p_T > 20$ GeV

 {\it e. No High Energy Jets:}  $E_{Jet} < 20$ GeV

 {\it f. No Missing Energy:}  {\rm Missing Transverse Energy} $< 15$ GeV

 {\it g. No High Energy or Isolated Leptons:} $p_{lepton} < 6$ GeV if
  not isolated
\begin{figure}
\resizebox{6.0in}{!}{
\includegraphics{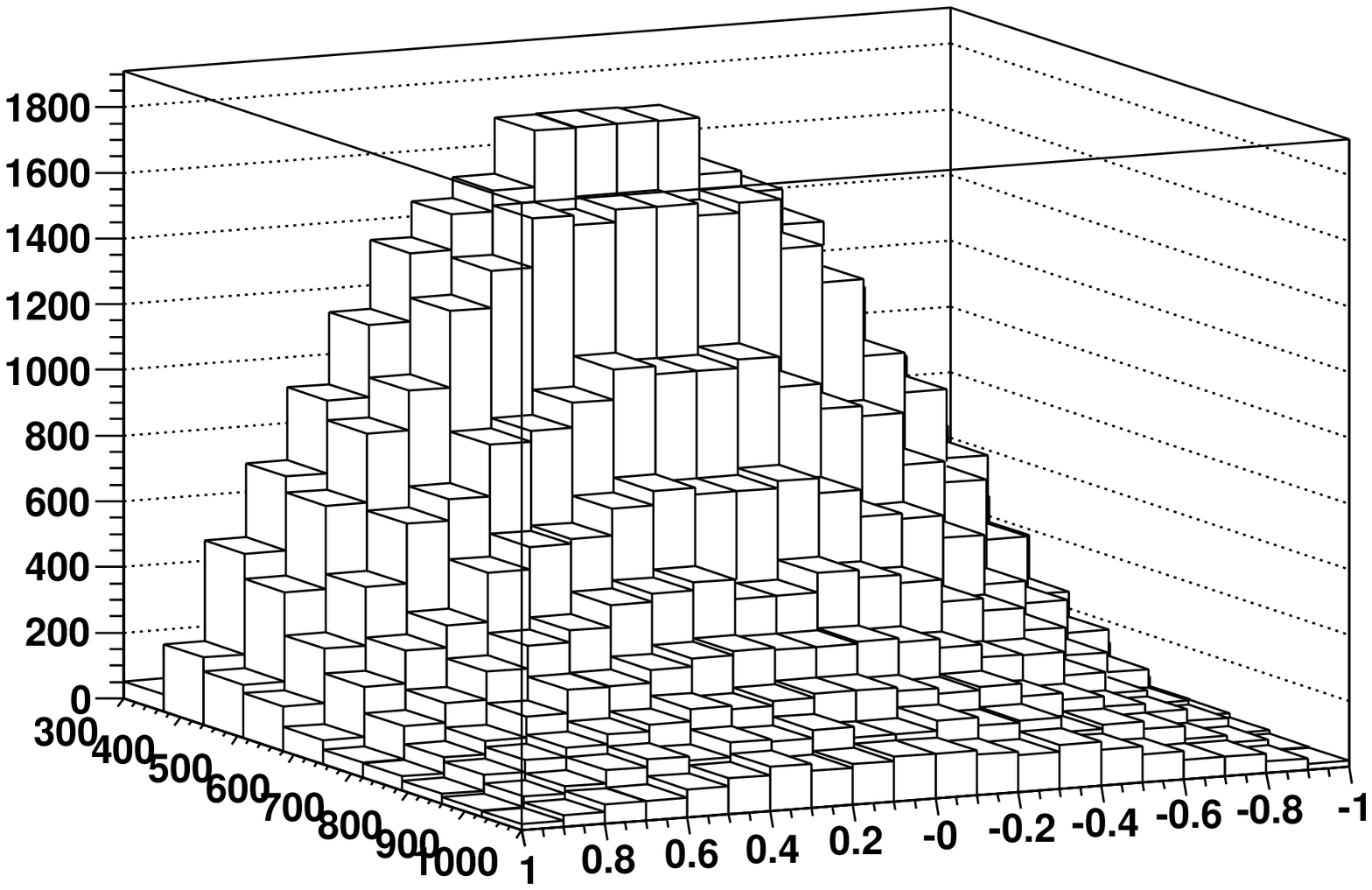} \qquad
\includegraphics{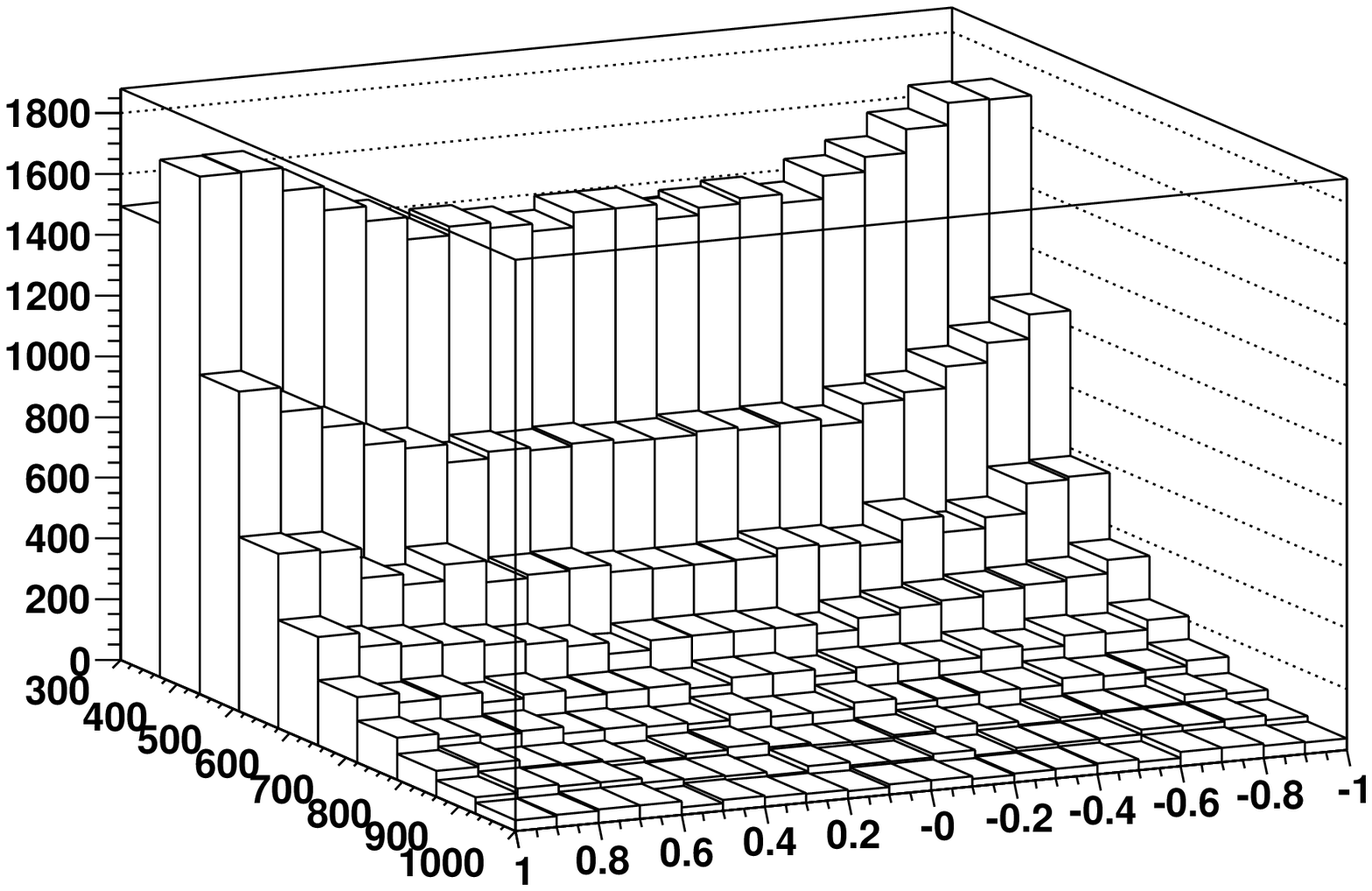}
} \caption{A 20$\times$20 two dimensional histogram of the center of
  mass kinematic variables for a 159 GeV scalar~(left) and
  fermion~(right).  The x-axis is cos$\theta^{*}$ with a range from -1
  to 1.  The y-axis is $\sqrt{s} = 2 E^{*}$ with a range from 300 GeV to 1000 GeV.
  \label{fig:double}}
\end{figure}

These cuts succeed at separating the sleptons from cascade decays
from the Drell-Yan produced particles so long as the particle
spectrum is well spaced, leading to large jet energies in cascade decays.
This type of spectrum is generic in mSugra space.

It should be noted that these cuts are very severe in general for the
extended mSugra space.  The cuts are meant to separate out the
sleptons produced from decays without knowing the mass spectrum, and
can be excessive for mass spectra like Model B where less than 1 out
of every 10000 slepton events that passes the cuts is not a Drell-Yan
event.  Less stringent cuts could be used with an increase in the
signal if the mass spectrum is taken into account.  On the other hand,
if the mass spectrum is more compressed than that of Model A, the
success of the cuts decreases.  These types of mass spectra are not
characteristic in the extended mSugra space, though they can be found
in other models. Specialized cuts for specific mass spectra would improve the
results.  This is not done here since the goal of this paper is to
determine a general and conservative prospect for spin determination
at the LHC.

After applying these cuts, the
effective Drell-Yan cross sections for the NLSP sleptons are 4.72 fb
for Model A and 1.05 fb for Model B.  The
cross sections for the sleptons from cascade decays are 0.187 fb for
Model A and less than 0.001 fb for
Model B.  The dual requirements of both slepton candidates having a
mass consistent with the measured slepton mass and a time delay
greater than 1 ns reduces the SM and BSM muon background to negligible
values.

We can therefore ignore the backgrounds in the further analysis, since they
are very small. We will assume that all events that pass the cuts above are
Drell Yan events, and we now proceed to an analysis of the spin measurement.

We note that the above cuts, which are necessary to remove these NLSP
sleptons not produced by Drell-Yan processes, alter the actual
distributions from naive theoretical calculations.
As a result, the experimental distribution must be compared to a theoretical
distribution with the same experimental cuts applied.

\section{Spin Determination}
\label{sec:spin}

  Scalars produced in Drell-Yan processes have a
$\textrm{sin}^2\theta^{*}$ angular distribution in the center of mass
frame 
independent
of the center of mass energy while Drell-Yan produced fermions have
a
$1+\left(\frac{E^{*2}-m^2}{E^{*2}+m^2}\right)\textrm{cos}^2\theta^{*}$
angular distribution in the center
of mass frame.  These distributions are shown in~\figref{double} for a
159.5 GeV scalar~(left) and fermion~(right).
We wish to determine how much luminosity is necessary to determine the
NLSP slepton's spin, and in particular, to establish that the angular distribution is consistent with a scalar rather than a fermion distribution.

The luminosity is effectively set by two numbers: The effective Drell-Yan
cross section after the cuts, and the number of Drell-Yan events
needed to distinguish between a spin-0 and spin-1/2 particle.
The average number of events needed for a spin determination is given by a
log ratio test for the two distributions \cite{orear}
\begin{equation}
\bar{N} = \frac{\textrm{Log}\left(R\right)}{\int \textrm{Log}\left(\frac{f_s\left(x\right)}{f_f\left(x\right)}\right)
  f_s\left(x\right) dx}.
\label{logratio}
\end{equation}
$R$ is the desired probability of having one distribution over the other
distribution for a given set of observables. We will require $R
= 10^3$,
which would imply that an experimental distribution is $10^3$ times
more likely to be a scalar
distribution than a fermion distribution.
The integral is over the entire observable space, parameterized by $E^{*}$ and
$\textrm{cos}~\theta^{*}$ in the present case, and the two functions, $f_s\left(x\right)$ and $f_f\left(x\right)$, are
the scalar and fermion probability distribution functions after the imposition of the cuts.

\begin{figure}
\resizebox{6.0in}{!}{
\includegraphics{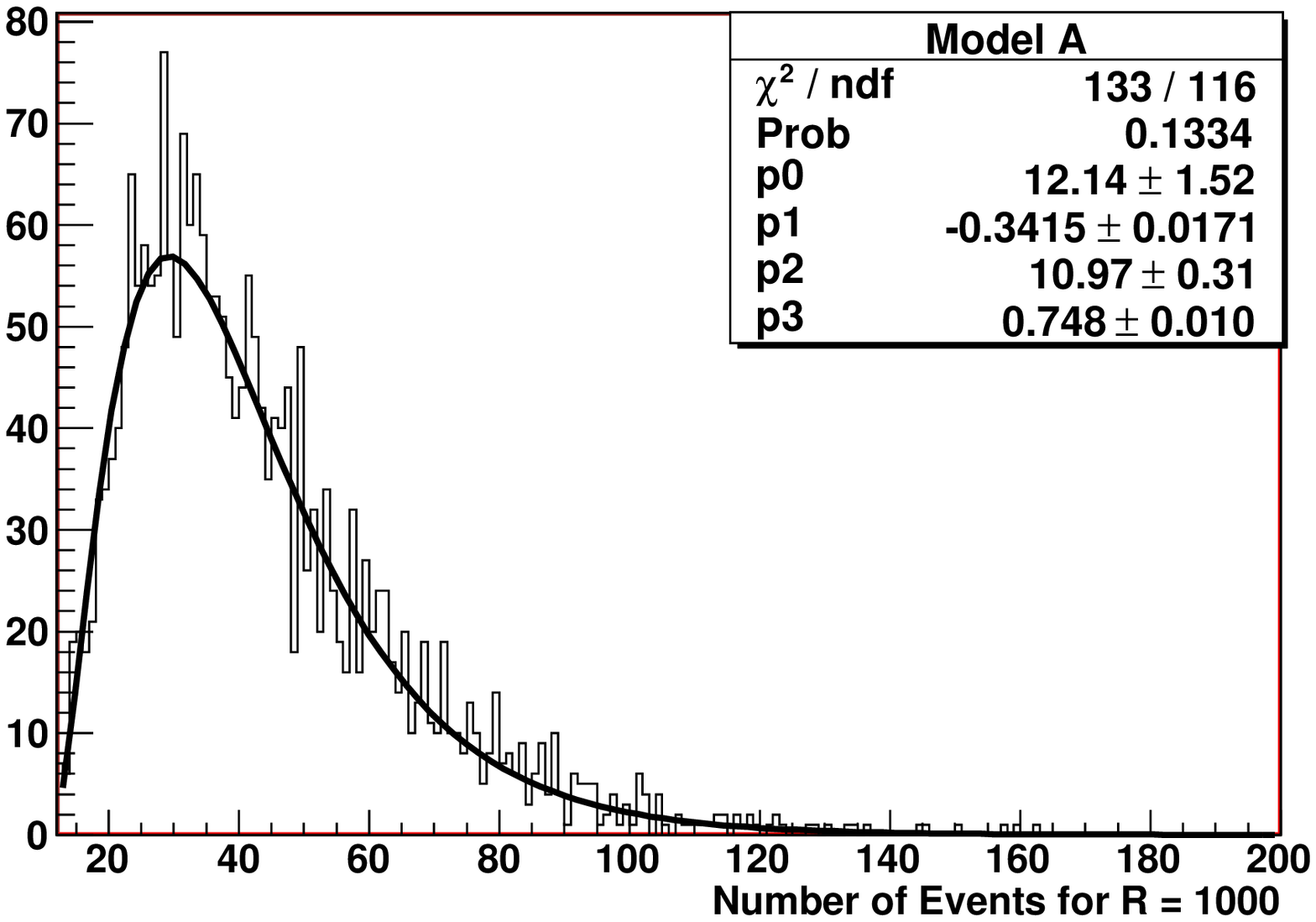} \qquad
\includegraphics{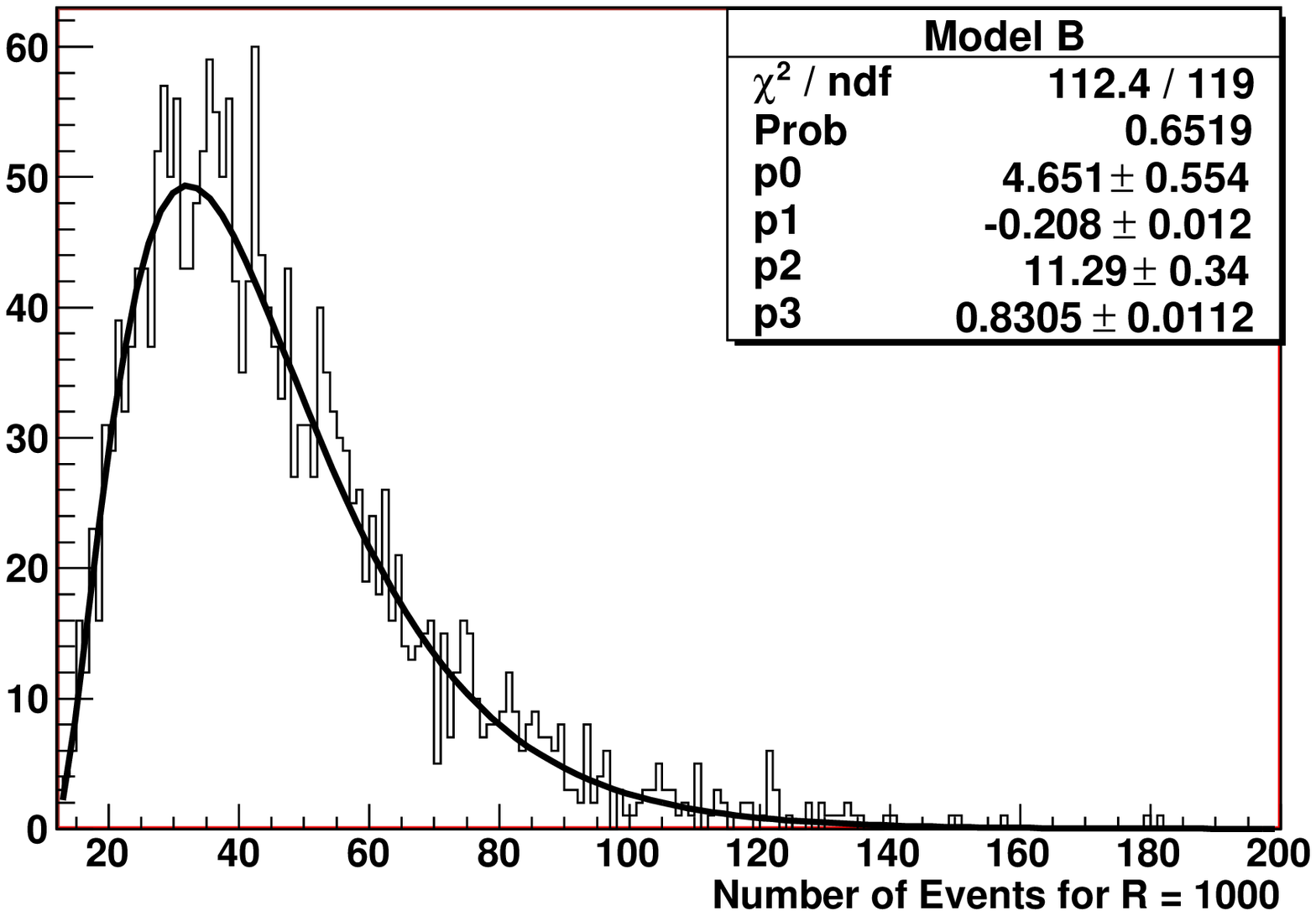}
} \caption{The histograms of the
number of events needed for R =
  1000.  
  The
  results are fitted well by~\eqref{fit} with the fitted parameters
  listed in the upper right hand corner of each panel.
\label{fig:pdf}}
\end{figure}

The probability distribution functions are generated using PYTHIA
6.406.
Drell-Yan events
were generated for a scalar and a
fermion with the measured mass of the model (including the previous uncertainties).  If the event passed the
cuts, it was then included into a 20$\times$20 two
dimensional ($E^{*}$,cos~$\theta^{*}$) grid until $10^5$ such events
were generated.  The large number of events is necessary to lower
the relative fluctuation per grid bin resulting in a  more accurate probability
distribution function.  The functions $f_s$ and $f_f$ are obtained by
normalizing the events in each grid bin.

The average number of events needed for a spin determination can then be calculated by
use of \eqref{logratio}. For the
two bench mark models, the numbers  are found to be
\begin{equation}
\bar{N}_A = \frac{\textrm{Log}\left(10^3\right)}{\sum_i \textrm{Log}\left(\frac{f^A_s\left(x_i\right)}{f^A_f\left(x_i\right)}\right)
  f^A_s\left(x_i\right)} \approx 41
\label{eqA}
\end{equation}
\begin{equation}
\bar{N}_B = \frac{\textrm{Log}\left(10^3\right)}{\sum_i \textrm{Log}\left(\frac{f^B_s\left(x_i\right)}{f^B_f\left(x_i\right)}\right)
  f^B_s\left(x_i\right)} \approx 44,
\label{eqB}
\end{equation}
and translate to average luminosities of $L_A = 8.7~
\rm{fb}^{-1}$ for Model A and $L_B = 42~\rm{fb}^{-1}$ for
Model B.  On average the spin of the slepton NLSP of Model A can be
measured by the end of the second LHC physics run.

 In general the number of events necessary to
distinguish between the scalars and fermions will fluctuate widely around
$\bar{N}$;  that is, over several different
runs, there is a large variation in the number of events required to get $R = 10^3$.
For any given luminosity, we should therefore provide a probability of
distinguishing between fermions and scalars.

The distributions of the number of events needed to determine the spin are shown in \figref{pdf}
for Model A~(left) and Model B~(right).
The distributions are constructed by generating  Drell-Yan events
for each model until $R \ge 10^3$.  The number of events needed
to reach this value of $R$ is then binned into a histogram until 2000 counts have
been made.

The distributions have a tail towards higher number of
events and are not fitted well by any common probability distribution function.
 They were fitted best
with a fitting function of the form
\begin{equation}
L\left(N\right) = p_0\left(N-p_2\right)~e^{p_1 N^{p_3}}
\label{fit}
\end{equation} with $p_0$ setting the overall scale of the
function, $p_1$ being the negative coefficient in the exponential,
$p_2$ setting the value which the function starts, and $p_3$ determining
with what power the exponential tails falls with. The best fit values for these
parameters can be found in
~\figref{pdf}.

The total probability of getting a
distribution capable of distinguishing between a fermion and a scalar
for a given number of events is then given by the integration of the
normalized probability distribution function in~\eqref{fit}, from the
cutoff to the given number of events.

\begin{figure}
\resizebox{3.0in}{!}{
\includegraphics{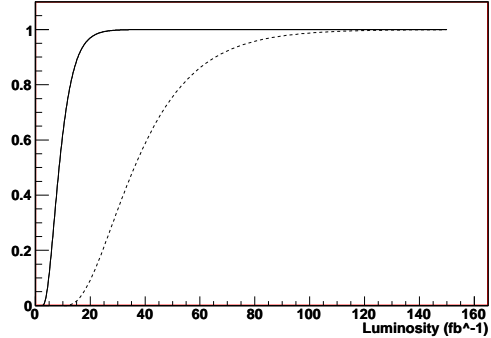} 
}
\caption{The  probability that the spin will be determined at the
  LHC as a
  function of luminosity.  The solid line is for the NLSP stau of Model
  A, and the dashed line is for the NLSP right handed selectron of
  Model B.   \label{fig:probpdf}}
\end{figure}


\begin{equation}
P\left(N\right) = C\int^N_{p_2} L\left(N^{'}\right) dN^{'}.
\label{prob}
\end{equation}
where $C$ is
a normalization factor. 

The probability $P(N)$
as a function of luminosity
is plotted in \figref{probpdf}.  It is constructed by using the
effective Drell-Yan cross
section of 4.72 fb for Model A and 1.05 fb for Model B.
By the end of the second 30 fb$^{-1}$
physics run the probability of getting a sample of
Drell-Yan events from which the spin can be determined are 99.8$\%$ for Model A and 34.6$\%$ for Model B.
The results differ for the two models because of the difference in the
Drell-Yan cross sections, an overall factor since $L\left(N\right)$
is a function of the number of events.

\begin{figure}
\resizebox{6.0in}{!}{
\includegraphics{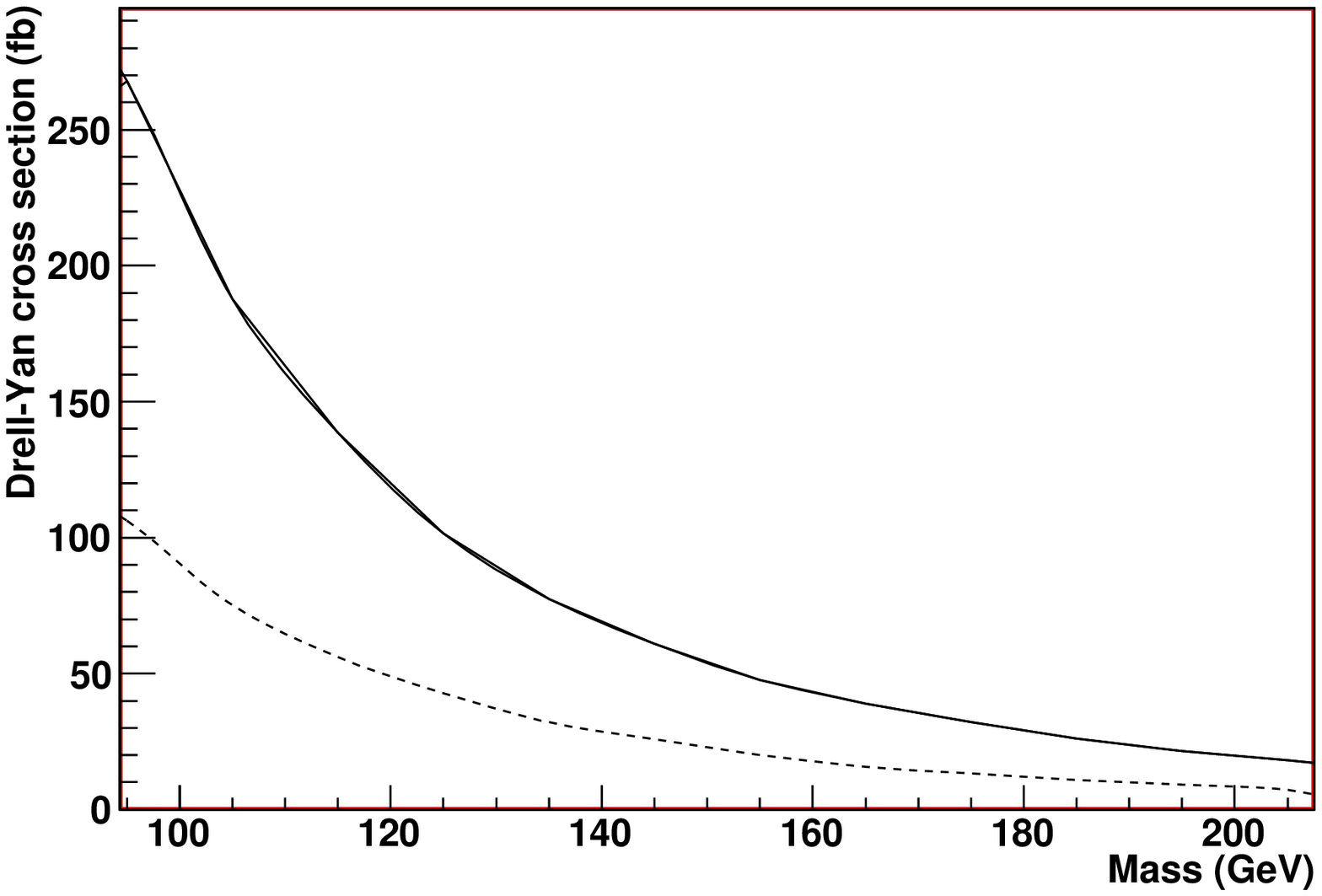} \qquad
\includegraphics{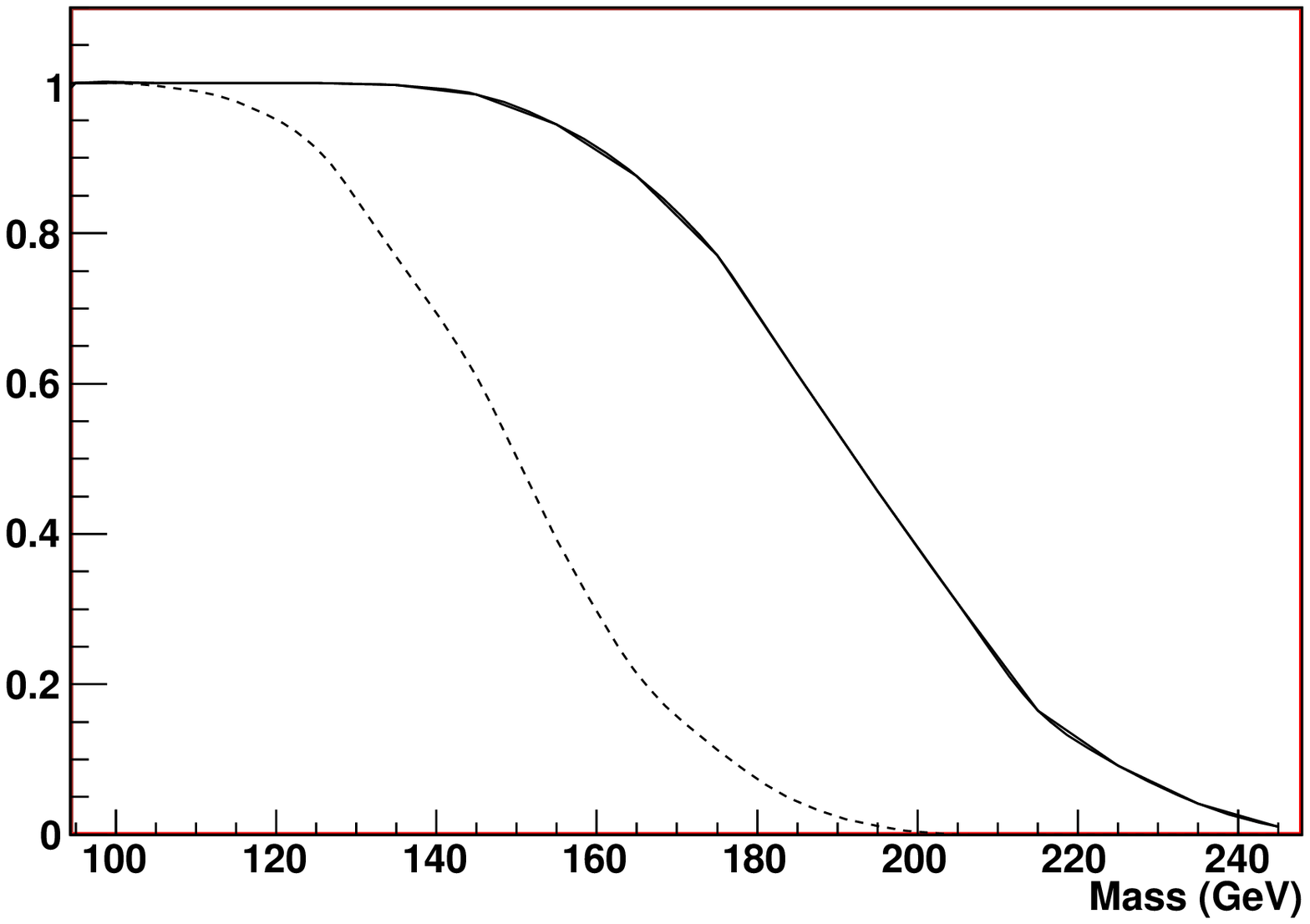}
}
\caption{The left panel is the slepton Drell-Yan cross section as a
  function of mass.  The right panel is
  the 
  probability that the spin will be determined
  by the end
  of the second 30 fb$^{-1}$ physics run as a function of slepton
  mass. In both panels the solid line is for a left handed slepton, and
  the dashed line represents a right handed slepton.  Results for mixed
  states lie between the two lines.  \label{fig:crossmass}}
\end{figure}

In general, the Drell-Yan cross section at
the LHC is a function of mass and of the amount of left and right
handedness in the mass eigenstate.
Using the results
from the fit combined with these Drell-Yan cross sections, an estimate
is made of the probability of producing a spin measurement by the end of the second LHC physics run as a function of
mass and handedness.  This is shown in ~\figref{crossmass}.
A spin determination
is $95\%$ probable for a left handed NLSP slepton with mass below 145 - 155
GeV.  Similarly, a right handed NLSP slepton spin
determination is probable for a mass below 115-125 GeV.

\section{Conclusion}
\label{sec:conclusion}
The first physics run of the LHC is expected to have a luminosity of
1-4 fb$^{-1}$.  For both 
benchmark models, the NLSP
slepton will be  discovered in this run.

The second physics run will
aim at acquiring 30 fb$^{-1}$ of luminosity.
In this run, we have shown that there is a good chance that the spin of
the slepton can be measured.
For Model A there is a 99.8$\%$ chance of determining the spin of
the 105 GeV stau, 
and for Model B there is
a 34.6$\%$ chance of determining the spin  of the 159 GeV right handed
selectron. On average these spin determinations
require 8.7 fb$^{-1}$ and 42 fb$^{-1}$ respectively for these two
models.

We have also estimated the chances of determining
the spin of the NLSP sleptons in a generic  scenario with metastable sleptons.  Slepton
masses less than 120 GeV will likely be verified as scalars by
the end of the second physics run, regardless of the handedness.  Left
handed sleptons can be shown to be scalar if the mass is less than 150
GeV.  Since most of the
mSugra parameter space with a NLSP slepton is largely right handed, it is
not likely that the spin of the slepton will be determined in the second physics run
at the LHC
if its mass is greater than 200 GeV.

Similar results can be derived for 100 fb$^{-1}$ and 300 fb$^{-1}$.
With 100 fb$^{-1}$, it is possible to determine the slepton spin  if the
mass is below 290 GeV, and with 300 fb$^{-1}$
the reach will extend to 380 GeV.

To conclude, we have shown that it is possible to measure the spin of
metastable charged particles
(e.g. metastable sleptons)
 after two years of LHC running.
It would be also be interesting to extend this analysis to see if the spins of the other
superpartners can be measured, using information from cascade decays
(for example, see~\cite{Barr:2005dz}). Such an analysis could
confirm the nature of the
new physics to be supersymmetry
while excluding extra dimension models.

\begin{acknowledgments}
The work of AR is supported in part by
NSF Grants No.~PHY--0354993 and PHY--0653656. BTS is supported in part by NSF CAREER
Grant No.~PHY--0239817.
\end{acknowledgments}


\begin{thebibliography}{99}

\bibitem{Cheng:2002ab}
  H.~C.~Cheng, K.~T.~Matchev and M.~Schmaltz,
  Phys.\ Rev.\  D {\bf 66}, 056006 (2002)
  [arXiv:hep-ph/0205314].

\bibitem{Barr:2004ze}
  A.~J.~Barr,
  Phys.\ Lett.\  B {\bf 596}, 205 (2004)
  [arXiv:hep-ph/0405052].

\bibitem{Smillie:2005ar}
  J.~M.~Smillie and B.~R.~Webber,
  JHEP {\bf 0510}, 069 (2005)
  [arXiv:hep-ph/0507170].

\bibitem{Datta:2005zs}
  A.~Datta, K.~Kong and K.~T.~Matchev,
  Phys.\ Rev.\  D {\bf 72}, 096006 (2005)
  [Erratum-ibid.\  D {\bf 72}, 119901 (2005)]
  [arXiv:hep-ph/0509246].

\bibitem{Barr:2005dz}
  A.~J.~Barr,
  JHEP {\bf 0602}, 042 (2006)
  [arXiv:hep-ph/0511115].

\bibitem{Alves:2006df}
  A.~Alves, O.~Eboli and T.~Plehn,
  Phys.\ Rev.\  D {\bf 74}, 095010 (2006)
  [arXiv:hep-ph/0605067].


\bibitem{Athanasiou:2006ef}
  C.~Athanasiou, C.~G.~Lester, J.~M.~Smillie and B.~R.~Webber,
  JHEP {\bf 0608}, 055 (2006)
  [arXiv:hep-ph/0605286].

\bibitem{Wang:2006hk}
  L.~T.~Wang and I.~Yavin,
  JHEP {\bf 0704}, 032 (2007)
  [arXiv:hep-ph/0605296].

\bibitem{Alves:2007xt}
  A.~Alves and O.~Eboli,
  arXiv:0704.0254 [hep-ph].


\bibitem{Chamseddine:1982jx}
  A.~H.~Chamseddine, R.~Arnowitt and P.~Nath,
  Phys.\ Rev.\ Lett.\  {\bf 49}, 970 (1982);
  R.~Barbieri, S.~Ferrara and C.~A.~Savoy,
  Phys.\ Lett.\ B {\bf 119}, 343 (1982);
  N.~Ohta,
  Prog.\ Theor.\ Phys.\  {\bf 70}, 542 (1983);
  L.~J.~Hall, J.~D.~Lykken and S.~Weinberg,
  Phys.\ Rev.\ D {\bf 27}, 2359 (1983);
  L.~Alvarez-Gaume, J.~Polchinski and M.~B.~Wise,
  Nucl.\ Phys.\ B {\bf 221}, 495 (1983).

\bibitem{Smith:1979rz}
  P.~F.~Smith and J.~R.~J.~Bennett,
  Nucl.\ Phys.\ B {\bf 149}, 525 (1979);
  P.~F.~Smith, J.~R.~J.~Bennett, G.~J.~Homer, J.~D.~Lewin,
H.~E.~Walford and W.~A.~Smith,
  Nucl.\ Phys.\ B {\bf 206}, 333 (1982);
  T.~K.~Hemmick {\it et al.},
  Phys.\ Rev.\ D {\bf 41}, 2074 (1990);
  P.~Verkerk, G.~Grynberg, B.~Pichard, M.~Spiro, S.~Zylberajch,
M.~E.~Goldberg and P.~Fayet,
  Phys.\ Rev.\ Lett.\  {\bf 68}, 1116 (1992);
  T.~Yamagata, Y.~Takamori and H.~Utsunomiya,
  Phys.\ Rev.\ D {\bf 47}, 1231 (1993).

\bibitem{Dimopoulos:1989hk}
  S.~Dimopoulos, D.~Eichler, R.~Esmailzadeh and G.~D.~Starkman,
  Phys.\ Rev.\ D {\bf 41}, 2388 (1990);
  M.~Byrne, C.~F.~Kolda and P.~Regan,
  Phys.\ Rev.\ D {\bf 66}, 075007 (2002)
  [hep-ph/0202252].

\bibitem{Pospelov:2006sc}
  M.~Pospelov,
  arXiv:hep-ph/0605215.

\bibitem{Cyburt:2006uv}
  R.~H.~Cyburt, J.~Ellis, B.~D.~Fields, K.~A.~Olive and V.~C.~Spanos,
  arXiv:astro-ph/0608562.


\bibitem{Feng:2005ba}
  J.~L.~Feng, A.~Rajaraman and B.~T.~Smith,
  Phys.\ Rev.\ D {\bf 74}, 015013 (2006)
  [arXiv:hep-ph/0512172].

\bibitem{Feng:2003xh}
J.~L.~Feng, A.~Rajaraman and F.~Takayama,
Phys.\ Rev.\ Lett.\  {\bf 91}, 011302 (2003) [hep-ph/0302215];
Phys.\ Rev.\ D {\bf 68}, 063504 (2003) [hep-ph/0306024].

\bibitem{Ellis:2003dn}
J.~R.~Ellis, K.~A.~Olive, Y.~Santoso and V.~C.~Spanos,
Phys.\ Lett.\ B {\bf 588}, 7 (2004) [hep-ph/0312262].

\bibitem{Feng:2004zu}
J.~L.~Feng, S.~f.~Su and F.~Takayama,
Phys.\ Rev.\ D {\bf 70}, 063514 (2004) [hep-ph/0404198];
Phys.\ Rev.\ D {\bf 70}, 075019 (2004) [hep-ph/0404231].

\bibitem{Wang:2004ib}
  F.~Wang and J.~M.~Yang,
  Eur.\ Phys.\ J.\ C {\bf 38}, 129 (2004)
  [hep-ph/0405186].

\bibitem{Ellis:2004bx}
  J.~R.~Ellis, K.~A.~Olive, Y.~Santoso and V.~C.~Spanos,
  Phys.\ Lett.\ B {\bf 603}, 51 (2004)
  [hep-ph/0408118].

\bibitem{Roszkowski:2004jd}
  L.~Roszkowski, R.~Ruiz de Austri and K.~Y.~Choi,
  JHEP {\bf 0508}, 080 (2005)
  [hep-ph/0408227].

\bibitem{Brandenburg:2005he}
  A.~Brandenburg, L.~Covi, K.~Hamaguchi, L.~Roszkowski and
  F.~D.~Steffen,
  Phys.\ Lett.\ B {\bf 617}, 99 (2005)
  [hep-ph/0501287].

\bibitem{Cerdeno:2005eu}
  D.~G.~Cerdeno, K.~Y.~Choi, K.~Jedamzik, L.~Roszkowski and R.~Ruiz de
  Austri,
  hep-ph/0509275.






\bibitem{Kaplinghat:2005sy}
M.~Kaplinghat,
Phys.\ Rev.\ D {\bf 72}, 063510 (2005) [astro-ph/0507300].

\bibitem{Cembranos:2005us}
J.~A.~R.~Cembranos, J.~L.~Feng, A.~Rajaraman and F.~Takayama,
Phys.\ Rev.\ Lett. {\bf 95}, 181301 (2005) [hep-ph/0507150].

\bibitem{Jedamzik:2005sx}
K.~Jedamzik, M.~Lemoine and G.~Moultaka,
astro-ph/0508141.

\bibitem{Sigurdson:2003vy}
  K.~Sigurdson and M.~Kamionkowski,
  Phys.\ Rev.\ Lett.\  {\bf 92}, 171302 (2004)
  [astro-ph/0311486].

\bibitem{Profumo:2004qt}
  S.~Profumo, K.~Sigurdson, P.~Ullio and M.~Kamionkowski,
  Phys.\ Rev.\ D {\bf 71}, 023518 (2005)
  [astro-ph/0410714].
  
  \bibitem{Steffen:2006hw}
   F.~D.~Steffen,
   JCAP {\bf 0609}, 001 (2006)
   [arXiv:hep-ph/0605306].
   
   \bibitem{Steffen:2006wx}
   F.~D.~Steffen,
   AIP Conf.\ Proc.\  {\bf 903}, 595 (2007)
   [arXiv:hep-ph/0611027].

\bibitem{Pradler:2006qh}
   J.~Pradler and F.~D.~Steffen,
   Phys.\ Rev.\  D {\bf 75}, 023509 (2007)
   [arXiv:hep-ph/0608344].

\bibitem{Rajaraman:2006mr}
  A.~Rajaraman and B.~T.~Smith,
  arXiv:hep-ph/0612235.



\bibitem{Ellis:2006vu}
  J.~R.~Ellis, A.~R.~Raklev and O.~K.~Oye,
  JHEP {\bf 0610}, 061 (2006)
  [arXiv:hep-ph/0607261].

\bibitem{Polesello:1999aa}
G.~Polesello, A.~Rimoldi, ``Reconstruction of quasi-stable charged
sleptons in the ATLAS Muon Spectrometer'', ATL-MUON-99-006.

\bibitem{orear}
J. Orear, Notes on Statistics for Physicists, UCRL-8417,
http://nedwww.ipac.caltech.edu/level5/Sept01/Orear/frames.html



\end{thebibliography}
\end{document}